# Thermodynamics and Relativity
## A Condensed Explanation of their Close Connection

Jean-Louis TANE, tanejl@aol.com
Formerly with the Department of Geology, University Joseph Fourier, Grenoble, France

**Abstract**: Although its practical efficiency is unquestionable, it is well known that thermodynamics presents conceptual difficulties from the theoretical point of view. It is shown that the problem comes from an imperfect compatibility between the first law and the second law. The solution suggested consists of connecting thermodynamics with relativity (and, by extension, with gravitation) through the Einstein mass-energy relation.

## 1. Introduction

The thermodynamic theory is mainly composed of two laws which have the character of postulates and are both very general and simple. To briefly recall what they are, let us consider a system defined as a given amount of gas which is placed in a vessel and whose exchanges with the near surroundings are limited to mechanical work and heat, excluding exchanges of matter. In thermodynamic langage, such a system is called a closed system. In contrast, it is sometimes useful to take into account the existence of a larger system defined as the sum gas, plus vessel, plus near surroundings, which is supposed to exchange neither energy nor matter with its own surroundings. This larger system is called an isolated system.

If the gas placed in the vessel evolves from an initial state $(P_1, V_1, T_1)$ to a final state $(P_2, V_2, T_2)$, the first law of thermodynamics relies on the idea that its change in internal energy is the same whether the process is reversible or irreversible.

Knowing that the change in internal energy, noted $\Delta U$ is the sum of the work exchanged, noted $\Delta W$, and of the heat exchanged, noted $\Delta Q$, this first law can be summarized through the equality $\Delta U_{irr} = \Delta U_{rev}$. Written in its differential form, it becomes:

$$dU_{irr} = dU_{rev} \qquad (1)$$

Correlatively, the differential form of the second law is usually presented under the expressions:

$$dQ_{rev} = TdS \qquad (2)$$

$$dQ_{irr} < TdS \qquad (3)$$

The exact meaning of expressions 2 and 3 will be seen further. It is enough to recall, for the moment, that dS represents the change in entropy, parameter whose link with the temperature T is comparable to that of the change of volume dV with the pressure P in expression:

$$dW = -PdV \qquad (4)$$

When they are connected together, the first and second laws of thermodynamics lead to the triplet of equations:

$$dU_{rev} = dQ_{rev} + dW_{rev} \qquad (5)$$
$$dU = TdS - PdV \qquad (6)$$
$$dU_{irr} = dQ_{irr} + dW_{irr} \qquad (7)$$

[with the fundamental postulate $dU_{irr} = dU_{rev}$]

The detailed explanation of these equations, which can be looked as "the basic thermodynamic tool" is given in many books on thermodynamics. The references quoted in this article are limited to some of them, chosen for their particular interest. Reference [1] is one of the most famous and exhaustive, reference [2] is a condensed book showing how the thermodynamic tool can be applied in chemistry, references [3] and [4] are more specially concerned with its application in geology, yet they present interesting comments from the theoretical point of view.

As will be shown below, there is a slight but fundamental inconsistency in our classical understanding and use of these equations.

## 2. Statement of the problem

It is well known, in physics, that the more general meaning of equation 4 is:

$$dW_{irr} = - P_e dV \qquad (8)$$

where $P_e$ is the pressure external to the system, i.e. the pressure of its near surroundings. If $P_e$ is equal to the internal pressure $P_i$, we are in conditions of reversibility and equation 8 becomes:

$$dW_{rev} = - P_i dV \qquad (9)$$

Referring to the gaseous system evoked above, let us imagine that, within the vessel, its upper separation from the near surroundings is a mobile piston of negligible weight. Whether the condition is $P_i > P_e$ or $P_i < P_e$, the volume of the gas will respectively increase ($dV > 0$) or decrease ($dV < 0$).

The volume being one of the parameters which define the state of the system, the value $dV$ is the same, for a given change of state, whether the process is reversible or irreversible. Thus, from eq. 8 and 9, we can deduce the relation:

$$dW_{irr} - dW_{rev} = dV( P_i - P_e) \qquad (10)$$

whose equivalent form is:

$$dW_{irr} = dW_{rev} + dV( P_i - P_e) \qquad (11)$$

Since $dV$ is positive when $P_i > P_e$ and negative when $P_i < P_e$ the term $dV( P_i - P_e)$ is always positive (except that it becomes zero in conditions of reversibility). Thus, the important information which needs to be memorized is:

$$dW_{irr} > dW_{rev} \qquad (12)$$

It has been recalled above that the general expression of the first law of thermodynamics is given by eq. 1, which states that $dU_{irr} = dU_{rev}$. To conciliate this data with that given by equation 12, the only possibility is that, in compensation, the relation between $dQ_{irr}$ and $dQ_{rev}$ be:

$$dQ_{irr} < dQ_{rev} \qquad (13)$$

At first glance, this last result seems to be in good accordance with equations 2 and 3 which, combined, lead effectively to the same proposition. Yet, this apparent coherency fails when we are confronted to the following situation.

Let us consider an isolated system consisting of a vessel divided in two parts separated by a mobile piston of negligible weight. We suppose that part 1 contains a gas whose pressure is $P_1$ and part 2 a gas whose pressure is $P_2$. If the piston, initially locked, is freed, it will move toward the field of lower pressure, so that the volume of part 1 varies of a quantity $\Delta V_1$, while the volume of part 2 varies of a quantity $\Delta V_2 = - \Delta V_1$.

This process is evidently irreversible and since the whole system is isolated, its internal energy remains constant and obeys the relation:

$$dU_{syst} = 0 \qquad (14)$$

Such a result means that, correlatively, the sum of the energetic exchanges between part 1 and part 2, must have a zero value. In order to see if this is true, we can proceed as follows.

According to eq. 8, the elementary changes in work for part 1, part 2 and the whole system are respectively:

$$dW_1 = - P_2 \, dV_1 \qquad (15)$$
$$dW_2 = - P_1 \, dV_2 \qquad (16)$$
$$dW_{syst} = dV_1 \, (P_1 - P_2) \qquad (17)$$

Since $dV_1$ is positive when $P_1 > P_2$ and negative when $P_1 < P_2$, the value of $dW_{syst}$, as already seen, is always positive.

Consequently, we have to verify that another exchange of energy between part 1 and part 2 has an opposite global value. Taking into account that the temperature increases in the compressed part and decreases in the expanded part, we easily conceive that this complementary exchange concerns heat. The problem is that, in the classical conception of thermodynamics, an exchange of heat between two bodies is exclusively understood as obeying the "law of heat exchange". This law states that if the first body reveives a heat $+ dQ_1$, the second body loses a heat $- dQ_2$ in such a way that we have:

$$dQ_{syst} = dQ_1 - dQ_2 = 0 \qquad (18)$$

For the elementary physical process which is considered presently, we don't see what other kind of energy (E) could have a global value $dE_{syst}$ which, added to $dW_{syst}$, lead to the expected conclusion:

$$dU_{syst} = 0$$

Thus the alternative conclusion is that we have reached a dead end whose exit needs a revision of our classical understanding of thermodynamics. Despite this situation, it is well known that the thermodynamic tool presents an indisputable efficiency in practice. This is certainly the sign that the problem is purely theoretical and requires, to be solved, that a detail generally unsuspected has to be taken into account.

Before examining this question, it can be useful to direct attention on the following point.

According to the convention adopted in thermodynamics, an energy is counted positively when it is received by the system and negatively when it is provided by the system. In the present article, this convention is strictly respected, but this is not always the case in scientific texts, where punctual exceptions can occur. The reason is that, for the pionniers of thermodynamics, the energy counted positively was the one available to the experimentator, i.e.

the one received by the near surroundings from the system. Among the books where this early convention is locally conserved is reference [1], where the information corresponding to equation 12 seems to be inverted. This impression is due to the fact that, on the graph giving a comparison between $dW_{irr}$ and $dW_{rev}$, the energetic quantity represented is not $-P\Delta V$ but $P\Delta V$. As a consequence, the inequality $dW_{irr} > dW_{rev}$ corresponding to eq. 12 appears under the form $dW_{irr} < dW_{rev}$ although its real meaning is $-dW_{irr} < -dW_{rev}$, i.e. $dW_{irr} > dW_{rev}$.

Keeping in mind that the understanding of the thermodynamic reasoning gets simplified when the convention of signs is systematically followed, we can now examine how the problem previously evoked can be solved.

## 3. Suggested solution

It can be easily seen that the difficulty encountered above disappears if we substitute the classical conception of the first law of thermodynamics by an extended one, i.e. if we substitute the postulate:

$$dU_{irr} = dU_{rev} \qquad (1)$$

by the postulate:

$$dU_{irr} > dU_{rev} \qquad (16)$$

For an isolated system, as just examined, the condition is $dU_{rev} = 0$ and $dU_{irr} > 0$.

Is it possible to conciliate the second law with this extended conception of the first law?

The answer is positive and can be argumented as follows. The second law is classically written under the form:

$$dS = dS_e + dS_i \qquad (19)$$

whose precise meaning is:

$$dS = \frac{dQ}{T_e} + dS_i \qquad (20)$$

Eq. 20 has the dimension of an entropy, but presented under the form:

$$T_e dS = dQ + T_e dS_i \qquad (21)$$

it takes the dimension of an energy and its meaning becomes:

$$dQ_{irr} = dQ_{rev} + dQ_{add} \qquad (22)$$

It is well-known in thermodynamics that the term $dS_i$ (which represents the internal component of entropy) has a positive value for an irreversible process and a zero value for a reversible process. Since $T_e$ is an absolute temperature, the proposition remains true for the energetic term $T_e dS_i$. Therefore the additional energy $dQ_{add}$ is positive and we can write:

$$dQ_{irr} > dQ_{re} \qquad (23)$$

An important point to be noted is the similarity between eq. 23 (which refers to the heat exchange) and eq. 12 (which refers to the work exchange).

Now, observing that eq. 16 can itself be written:

$$dU_{irr} = dU_{rev} + dU_{add} \qquad (24)$$

where $dU_{add}$ is a positive term, we see that there is a close analogy between eq. 24 and eq. 21, that is between the extended expression of the first law (eq. 24) and the extended expression of the second law (eq. 21).

Of course, if the substitution of eq. 1 by eq. 16 offers a possibility, from the theoretical point of view, to eliminate the inconsistency evoked above, the problem remains, from the practical point of view, to explain how the term $dS_i$ (eq. 20) can be converted into the term $T_e dS_i$ (eq. 21). In the first case, $dS_i$ has the dimension of an entropy and the idea that its value can go increasing has been accepted for a long time by scientists. In the second case, $T_e dS_i$ has the dimension of an energy and the idea has never been envisaged, in thermodynamics, that an energy could be created within a system. In this field of physics, the energies taken into account are exclusively linked to energetic exchanges between the system which is considered and its surroundings.

The answer to this fondamental question can be exposed in three points.

1) When the laws of thermodynamics have been stated in the XIXth century, it was quite impossible for their creators to introduce in their reasoning the idea that an energy could be created within a system. At that time, one of the main principles of physics was the law of conservation of energy to which any new theory was necessarily submitted.

2) The situation is different presently since, in the meantime, the possibility has been revealed by Einstein that, within a system, an increase in energy can be generated by a decrease in mass and conversely. The law of conservation of energy is maintained, but its signification is extended.

It is evidently the argument invoked in this paper to explain the origin of the term $T_e dS_i$ in eq. 21 or of the term $dU_{add}$ in eq. 24. Recalling that the mass-energy relation is $E = mc^2$, its differential form is $dE = c^2 dm$, which can be better written $dE = -c^2 dm$, in the present case, to respect the thermodynamic convention of signs previously evoked.

As a consequence, the term $T_e dS_i$ in eq. 21 as well as the term $dU_{add}$ in eq. 24 are considered as having the signification and the value $dE = -c^2 dm$ of the Einstein relation.

3) As recalled above, there is an analogy between expression $dW = -PdV$ and expression $dQ = TdS$. Let us come back to the experimental context of a vessel containing a gas topped by a mobile piston of negligible weight. If the pressure $P_e$ is greater that $P_i$, the decrease in volume whose value, according to eq. 8, is $dV = -dW/P_e$, appears as the sign that the system receives work. The only condition to conclude that the system does not receive (or provide) work would be the absence of change in volume.

The situation is the same for the link between energy and entropy in relation $dQ = TdS$. A change in entropy is the sign of a change in energy and when the change in entropy concerns the internal component $dS_i$, the corresponding change in energy concerns the internal component $dU_i$. The reason why this connection has not been looked as an evidence is certainly due to the fact that, contrary to a change in volume, a change in entropy is not a process that we can perceive directly.

## 4. Condensed presentation of the "extended thermodynamic tool"

In the same manner as the "classical thermodynamic tool"can be summarized through the set of equations 5, 6 and 7 (to which is added the fundamental postulate $d_{irr} = d_{rev}$), the "extended thermodynamic tool"can be summarized through the following set of equations:

$$dU_{rev} = dQ_{rev} + dW_{rev} \qquad (25)$$
$$dU_{rev} = T_i dS - P_i dV \qquad (26)$$
$$dU_{irr} = dQ_{irr} + dW_{irr} \qquad (27)$$
$$dU_{irr} = T_e dS - P_e dV \qquad (28)$$

[with the fundamental postulate $d_{irr} > d_{rev}$]

In this extended conception, where thermodynamics and relativity are closely connected, the fundamental equation takes the form:

$$dU_{irr} = dU_{rev} - c^2 dm \qquad (29)$$

where dm has a negative value.

Compared with the classical conception, the novelty which is introduced lies in the idea that the law of evolution of a thermodynamic system, usually interpreted as an increase in entropy, is interpreted as an increase in energy, linked to a correlative decrease in mass.

For a better understanding of the difference between the two interpretations, two examples are examined below. A first example deals with the expansion of a gas into vacuum, which can be looked as a particular case of exchange of work between two parts of an isolated system. A second example is devoted to an exchange of heat.

## 5. Exchange of work between two parts of an isolated system

Let us come back to the experimental context of an isolated system consisting of a vessel divided in two parts separated by a mobile piston of negligible weight. As already done, we suppose that part 1 contains a gas whose pressure is $P_1$ and part 2 a gas whose pressure is $P_2$. We have seen above that the elementary changes in work for part 1, part 2 and the whole system are respectively:

$$dW_1 = -P_2 dV_1 \qquad (15)$$
$$dW_2 = -P_1 dV_2 \qquad (16)$$
$$dW_{syst} = dV_1 (P_1 - P_2) \qquad (17)$$

As already noted, the value of $dW_{syst}$, in eq. 17, is necessarily positive.

Confronted with this situation, the thermodynamic analysis of the expansion of a gas into vacuum is not the same, depending upon whether we adopt the classical interpretation or the one suggested presently.

**5.1. Classical interpretation**. The thermodynamic tool which is used is the set of equations 5, 6 and 7, associated with the postulate $dU_{irr} = dU_{rev}$. Keeping in mind that the vacuum has no energetic effect, we focus attention exclusively on the gas, i.e. on part 1.

To calculate the work, we refer to the general equation $dW = -P_e dV$ and having $P_e = 0$ ($P_e$ is the pressure $P_2$ of the vacuum) we write $dW = 0$. Concerning the heat exchange, we write

dQ = 0 (since the gas cannot exchange heat with the vacuum). The global energetic result is thus $dU_{syst} = 0$, a proposition which is looked as being in accordance with the first law of thermodynamics. Then, taking into account eq. 6, we get dS = P/TdV, and admitting that all the parameters now refer to the gas (an assumption whose logic is not evident) we conclude that $dS_{syst}$ has a positive value. This last result is looked as being in accordance with the second law of thermodynamics.

**5.2. Extended interpretation**. It consists of applying equation $dW = - P_e dV$ to each part of the system, the gas and the vacuum, as was already done with eq. 15 and 16. Therefore, the global work exchange is obtained by entering $P_2 = 0$ in eq. 17 and takes the form:

$$dW_{syst} = dV_1 \, P_1 \qquad (30)$$

whose value is positive.

Then, having dQ = 0 (for the same reason as that evoked in the classical interpretation), we obtain for dU:

$$dU_{syst} = P_1 \, dV_1 \qquad (31)$$

whose value is thus positive.

This result is interpreted as an increase in energy linked to a correlative decrease in mass, according to the Einstein mass-energy relation.

Referring to this example, which concerns an exchange of work, one the major differences between the classical interpretation and the extended one lies in the fact that, in the latter, the concept of increase in entropy has not been taken into account. It has been directly susbstituted by the concept of increase in energy and its natural correlation to a decrease in mass.

The concept of entropy being closely connected to the link between heat and temperature, it is interesting to examine now, for an exchange of heat, what the difference is between the classical and the extended interpretation.

## 6. Exchange of heat between two parts of an isolated system

This aspect of the problem is more easily illustrated through a numerical example.

Let us consider an isolated system consisting of a vessel which is composed of two parts. We suppose that part 1 contains a definite mass of water, $m_1$, whose initial temperature is $T_1 = 293$ K and part 2 a definite mass of water, $m_2$, whose initial temperature is $T_2 = 333$ K. The average heat capacity of water being c = 4184 J.kg$^{-1}$.K$^{-1}$, we can simplify the calculation (without restraining its interest) by choosing the same round value 1000 J.K$^{-1}$ for each of the global heat capacities, i.e. for $C_1$ (= $m_1 c$) and $C_2$ (= $m_2 c$). This amounts to give to $m_1$ and $m_2$ the commun value 0.239 kg. Since the water is always liquid in the present case, c corresponds indifferently to $c_p$ or $c_v$, whose values are practically the same.

We know that the natural evolution of such a system results in an irreversible exchange of heat between part 1 and part 2 until they reach the same final temperature $T_f$. This temperature can be calculated (and therefore predicted) using equation:

$$T_f = \frac{C_1 T_1 + C_2 T_2}{C_1 + C_2} \qquad (32)$$

In the present case we get $T_f = 313$ K

Admitting that there is no change in volume, and that this proposition is true not only for the whole system (defined as isolated), but also for part 1 and part 2, all the values $\Delta W_1$, $\Delta W_2$ and $\Delta W_{syst}$ are zero. Thus, referring to the classical thermodynamic tool (triplet of equations 5, 6 and 7), we get for each part of the system the condition:

$$\Delta U = \Delta Q \tag{33}$$

Beyond these considerations, the thermodynamic interpretation of the process is not the same, depending on whether we admit the classical conception or the new suggested one. The difference can be summarized as follows.

**6.1. Classical interpretation**. For each part of the system, the temperature evolves from the initial value $T_i$ to the final common value $T_f$ and the exchange of heat is given by the general equation $\Delta Q = C (T_f - T_i)$. Thus, we get:

$$\Delta Q_1 = 1000 (313 - 293) = 20000 \text{ J}$$
$$\Delta Q_2 = 1000 (313 - 333) = -20000 \text{ J}$$
$$\Delta Q_{syst} = \Delta Q_1 + \Delta Q_2 = 0$$
$$\Delta U_{syst} = \Delta Q_{syst} = 0$$

Being zero, this last value is looked as being in accordance with the classical conception of the first law of thermodynamics, since the whole system we are considering is isolated.

For the changes in entropy the general equation is:

$$\Delta S = \int_{T_i}^{T_f} \frac{C}{T} dT \tag{34}$$

If we consider that C does not vary significantly with T, eq. 34 can be substituted by:

$$\Delta S = C \, Ln \frac{T_f}{T_i} \tag{35}$$

and we infer respectively:

$$\Delta S_1 = 1000 \, Ln \, (313/293) = 66.03 \text{ J.K}^{-1}$$
$$\Delta S_2 = 1000 \, Ln \, (313/333) = -61.94 \text{ J.K}^{-1}$$
$$\Delta S_{syst} = \Delta S_1 + \Delta S_2 = 4.09 \text{ J.K}^{-1}$$

Being positive, this last value is looked as being in accordance with the classical conception of the second law of thermodynamics.

**6.2. Extended interpretation.** We have seen above that eq. 20, which has the dimension of an entropy, needs to be substituted by eq. 21 which has the dimension of an energy and corresponds to the differential expression:

$$T_e dS = dQ + T_e dS_i \tag{21}$$

We have also seen that eq. 21 is a particular case of eq. 24, whose differential expression is:

$$dU_{irr} = dU_{rev} + dU_{add} \tag{22}$$

where $dU_{add}$ is a positive term

By integration, eq. 21 and 24 take the forms:

$$T_e^* \Delta S = \Delta Q + T_e^* \Delta S_i \qquad (36)$$

$$\Delta U_{irr} = \Delta U_{rev} + \Delta U_{add} \qquad (37)$$

where $T_e^*$ is the average value of the external temperature $T_e$ during the process, i.e. the average value of $T_2$ for part 1 and the average value of $T_1$ for part 2. We can label them $T_2^*$ and $T_1^*$ and consider that they are **space-time parameters**, in the sense that each of them represents a temperature which is varying both in time and space. The same remark stands for P when we have to integrate expressions such as eq. 15, 16 and 17.

The numerical values $T_1^*$ and $T_2^*$ can be calculated by entering the preliminary results obtained above in the general relation $T = \Delta Q/\Delta S$. We get, in the present case:

$$T_1^* = \Delta Q_1/\Delta S_1 = 20000/66.03 = 302.89 \text{ K}$$
$$T_2^* = \Delta Q_2/\Delta S_2 = -20000/-61.94 = 322.89 \text{ K}$$

Now, entering these values in eq. 36, we get:

For part 1: $\quad T_2^* \Delta S_1 = \Delta Q_1 + T_2^* \Delta S_{i1}$
i.e. $\quad 21320 = 20000 + 1320$
whose significance is: $\quad \Delta U_{irr1} = \Delta U_{rev1} + \Delta U_{add1}$

For part 2: $\quad T_1^* \Delta S_2 = \Delta Q_2 + T_2^* \Delta S_{i2}$
i.e. $\quad -18760 = -20000 + 1240$
whose significance is: $\quad \Delta U_{irr2} = \Delta U_{rev2} + \Delta U_{add2}$

Adding these results, we obtain for the whole system:

$$2560 = 0 + 2560$$
whose significance is: $\quad \Delta U_{irr.syst} = \Delta U_{rev.syst} + \Delta U_{add.syst} \qquad (38)$

Although it does not appear directly, an interesting point to be observed, which is a general property of a heat exchange, is the equality $\Delta S_{i1} = \Delta S_{i2}$. Indeed, we have:

$$\Delta S_{i1} = 1320/T_2^* = 1320/322.89 = 4.09 \text{ J.K}^{-1}$$
$$\Delta S_{i2} = 1320/T_1^* = 1240/302.89 = 4.09 \text{ J.K}^{-1}$$

Therefore, when summing the energetic contributions of part 1 and part 2, the term $\Delta U_{add.syst}$ of eq. 38 can also be written:

$$\Delta U_{add.syst} = \Delta S_{i1} (T_2^* + T_1^*)$$
i.e. $\Delta U_{add.syst} = 4.09 (322.89 + 302.89) = 2560$

This value is evidently the same as the one previously obtained, but this altenative way shows that, in the calculation of a heat exchange, the term $\Delta S_{i1}$ can be factorized, in the same manner as was done for $\Delta V_1$ (whose meaning is $\Delta V_{i1}$) in the above calculation of a work exchange.

Now, remembering that the value 2560 is given in Joule and represents an increase in energy, the corresponding decrease in mass, according to the Einstein mass-energy relation, is:

$$\Delta m = -2560/(3 \times 10^8)^2 = -2.84 \; 10^{-14} \text{ kg} \qquad (39)$$

This decrease in mass is too small to be experimentally detectable. Yet, it provides an explanation to the correlative increase in energy which has been recognized itself as a necessary condition to make the laws of thermodynamics compatible.

## 7. Conclusion

Although the practical efficiency of thermodynamics is not contestable, it is well-known that its understanding raises difficulties which are specific to this field of physics. It has been shown above that the compatibility between the the first and second laws of thermodynamics is imperfect when they are taken in their usual conception, so that the problem is real. The suggested solution, whose synthetic expression is equation 29, consists of inserting the Einstein mass-energy relation in the usual theory, therefore combining thermodynamics and relativity. If recognized valid, this suggestion could contribute to further openings such as a simplification in the teaching of thermodynamics and the search of links between thermodynamics and gravitation. The link with gravitation is evidenced by the presence of mass (in the relativistic sense of this concept) in equation 29. In our near surroundings, the evolution of a system is considered normal when it results in a decrease in mass (extension of the classical idea of an increase in entropy). It is not excluded that in other systems, such as black holes, the rule is inverted and, on Earth, we are not sure that the thermodynamic evolution goes in the same direction for both inert matter and living matter.

The conceptual difficulties of thermodynamics are commented, more or less in detail, in references [1 to 4] quoted below. About this topic, an interesting confession from Arnold Sommerfeld can be found in the preface of reference [4]. Concerning the possibility of applying the extended conception suggested here to chemical reactions, some preliminary examples are evoked in reference [5].

**Acknowledgements**. I would like to thank the Journal of Theoretics for having accepted to publish my first papers on this subject.